\documentstyle[multicol,aps,pre,epsf,amstex]{revtex}

\begin{document}

\title{Coarsening in a Driven Ising Chain with Conserved Dynamics}
\author{V. Spirin, P. L. Krapivsky, and S. Redner}
\address{Center for BioDynamics, Center for Polymer Studies, 
and Department of Physics, Boston University, Boston, MA 02215}

\maketitle

\begin{abstract}

\noindent  
We study the low-temperature coarsening of an Ising chain subject to
spin-exchange dynamics and a small driving force.  This dynamical system
reduces to a domain diffusion process, in which entire domains undergo
nearest-neighbor hopping, except for the shortest domains -- dimers -- which
undergo long-range hopping.  This system is characterized by {\em two}
independent length scales: the average domain length $L(t)\sim t^{1/2}$ and
the average dimer hopping distance $\ell(t)\sim \sqrt{L(t)}\sim t^{1/4}$.  As
a consequence of these two scales, the density $C_k(t)$ of domains of length
$k$ does not obey scaling.  This breakdown of scaling also leads to the
density of short domains decaying as $t^{-5/4}$, instead of the $t^{-3/2}$
decay that would arise from pure domain diffusion.

\end{abstract}

\begin{multicols}{2}

\section{Introduction}

The approach to equilibrium in isotropic systems which are quenched from a
high-temperature homogeneous phase to a low-temperature two-phase region is
now relatively well understood\cite{gunton,ajb}.  The basic feature of such
systems is that they typically organize into a coarsening mosaic of
single-phase regions, with a characteristic length scale that grows as a
power law in time.  For driven systems, on the other hand, considerably less
progress has been made in understanding the coarsening dynamics, although the
stationary properties have been thoroughly investigated \cite{schm}.  In the
presence of driving, the physically relevant coarsening mechanisms are those
with conserved order-parameter dynamics.  This would be appropriate, for
example, for treating the phase separation of binary liquids or binary alloys
under the influence of gravity \cite{siggia}.

In this spirit, Cornell and Bray \cite {cobra} recently studied the
coarsening dynamics of an Ising chain which is endowed with conserved
spin-exchange Kawasaki dynamics and which is also subject to a driving field
which favors transport of up spins to the right and down spins to the left.
They argue that in the limit of low temperature and weak field, the spin
dynamics is equivalent to domain diffusion (DD) \cite{cobra}.  In this DD
process, up domains hop rigidly by one lattice spacing to the right, and down
domains hop by one lattice spacing to the left.  Due to this nearest-neighbor
hopping, small domains are progressively ``squeezed out'' and the adjacent
neighboring domains coalesce.  This random walk mechanism leads both to a
reduction in the number of domains and an increase in their average length,
$L(t)\sim t^{1/2}$.  Numerical evidence was also presented that the density
of domains of length $k$ obeys the scaling form $C_k(t)\sim
(k/L^3)\exp(-k^2/L^2)$ \cite{cobra}.  This further implies a $t^{-3/2}$
asymptotic decay for the density of domains of fixed length.

The goal of this paper is to show that there is a subtle but crucial
difference between the dynamics of individual spins in the driven Ising chain
with spin conserving dynamics and the DD process.  The fundamental point is
that for the shortest domains -- dimers -- the spin-level dynamics results in
{\it long-range} dimer hopping, with the average jump length growing as
$\sqrt{L}$.  In contrast, for the DD process, dimers necessarily jump to the
next domain boundary, that is, they jump a distance of order $L$.  Therefore
dimers disappear more slowly than in the DD process, and the overall
coarsening rate is dimer controlled (DC).  This DC system is characterized by
two independent length scales: (a) the average domain length $L(t)$, which is
still proportional to $t^{1/2}$, as in the DD process, and (b) the average
dimer hopping distance, which is proportional to $t^{1/4}$.  As a result of
the two length scales, the density of domains of fixed length asymptotically
decays as $t^{-5/4}$, instead of the $t^{-3/2}$ decay of the DD process.
Correspondingly, the domain length distribution in the DC process does not
obey conventional scaling in the small-length limit, although the domain
length distributions for both the DD and DC processes are visually similar.

In Sec.\ II, we define the spin dynamics precisely and describe the
correspondences to the DD and DC processes.  We also discuss their essential
differences and show how the latter process provides a faithful description
of the microscopic spin dynamics.  We then present heuristic arguments which
suggest new kinetic behavior for the DC process.  Simulation results which
support our basic arguments are presented in Sec.\ III.  In Sec.\ IV, we
outline a perturbative approach, based on a matched asymptotic expansion,
which accounts for the observed breakdown of scaling in the domain length
distribution for vanishingly small minority fraction.  Sec.\ V contains both
a summary and a brief discussion of open issues.

\section{Geometrical picture of the dynamics}

The microscopic system is a chain of Ising spins with nearest-neighbor
ferromagnetic interaction $J$.  The chain is subject to spin-exchange
dynamics, where the only possible re-arrangement process is the exchange of
two anti-parallel nearest-neighbor spins. Thus the magnetization is
manifestly conserved (we use a magnetic terminology, although a system with
conserved dynamics naturally applies to an alloy).  The exchange occurs at a
rate proportional to $e^{-\Delta/T}$, where $\Delta$ is the energy difference
between the initial and final states, and $T$ is the temperature (with
Boltzmann constant set to unity).  There is also a driving field $E$ which
favors motion of up spins to the right and down spins to the left.  The
spin-flip events are:
\begin{equation*}
%\label{proc}
\begin{array}{lll}
++-- \,\,\rightleftharpoons \,\,+-+- &\quad \Delta=4J-E &\quad (\hbox{i}),\\
--++ \,\,\rightleftharpoons \,\,-+-+ &\quad \Delta=4J+E &\quad (\hbox{ii}),\\
++-+ \,\,\rightleftharpoons \,\,+-++ &\quad \Delta=-E   &\quad  (\hbox{iii}),\\
-+-- \,\,\rightleftharpoons \,\,--+- &\quad \Delta=-E   &\quad  (\hbox{iv}).
\end{array}
\end{equation*} 
The first two processes occur on domain boundaries, while the last two account
for the motion of a single spin which is inside a domain of the opposite
sign.  The ``forward'' processes involve the energy change $\Delta$, while
the ``backward'' processes have energy change $-\Delta$.  

Interesting dynamics arises in the limit of low (but non-zero) temperature
and weak driving field, that is, $0<T\ll E\ll J$.  To appreciate the nature
of the dynamics for this parameter range, notice that in one dimension (1D)
the order-disorder transition occurs at $T_c=0$.  At $T=0$, the spin-exchange
dynamics traps the system in a metastable state which consists of domains of
lengths $\geq 2$ \cite{zero,cor}.  To avoid this ``freezing'', the
temperature must be non-zero.  At low but non-zero temperature, the system
will coarsen as long as the mean domain length is smaller than the
correlation length $\xi\sim e^{J/T}\gg 1$.

The limit where the driving field satisfies $T\ll E\ll J$ leads to an {\em
  approximate} equivalence with the DD process \cite{cobra}.  To understand
this correspondence, consider the situation after domains have coarsened to
relatively large length.  By process $({\rm i}\rightharpoonup)$, an up spin
may detach from the right edge of an up domain with rate $e^{-(4J-E)/T}$, or
equivalently, a down spin may detach from the left edge of a down domain.
Similarly, an up spin may also detach from the left edge of an up domain (or
a down spin may detach from the right edge of down domain) by step
$(\hbox{ii}\rightharpoonup)$.  However, this process occurs at a rate which
is a factor $e^{-2E/T}$ smaller than $({\rm i}\rightharpoonup)$.  Moreover,
even if step $({\rm ii}\rightharpoonup)$ occurs, the detached spin quickly
recombines with the same domain by the reverse process $({\rm
  ii}\leftharpoondown)$, since the motion of the detached spin away from the
domain is energetically unfavorable.

Once $(\hbox{i}\rightharpoonup)$ occurs, the system evolves further either by
$(\hbox{iv}\rightharpoonup)$, which corresponds to the up spin moving to the
right and eventually joining the next up domain, or by
$(\hbox{iii}\rightharpoonup)$, where a down spin moves to the left and joins
the next down domain.  The former process is illustrated in Fig.~1.  As a
result of these processes, an up domain hops rigidly by one lattice spacing
to the right or a down domain hops one spacing to the left.  If we measure
the time in units of $e^{(4J-E)/T}$, then both these hopping steps occur at
unit rate, {\em independent\/} of the domain length.

\begin{figure}
\narrowtext
\epsfxsize=5.6cm\epsfysize=4.0cm
\hskip 0.5in\epsfbox{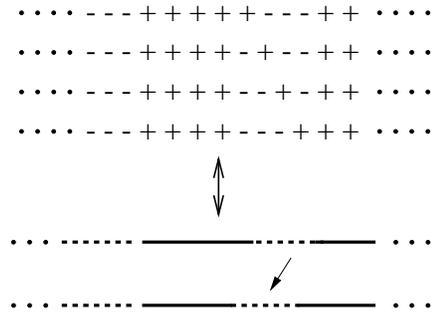}
\vskip 0.15in
\caption{Illustration of the detachment of an up spin from an up domain and 
  its merging with the neighboring up domain to the right.  In the upper part
  of the figure, each line represents the state of the system after a single
  spin-exchange event.  This evolution is equivalent to rigid-body domain
  hopping, with a down domain hopping one lattice site to the left, as
  indicated in the lower portion.
\label{fig1}}
\end{figure}

An essential feature of the low-temperature weak-field limit $T\ll E\ll J$ is
that all other processes are asymptotically negligible in the
intermediate-time range where coarsening is occurring.  Thus in this time
range the system consists of the contiguous array of alternating up and down
domains, and the dynamics proceeds by taking an up (down) domain and moving
it one lattice site to the right (left).  Whenever a domain shrinks to zero
size, its two adjacent neighbors coalesce.  This description is the basis of
the correspondence to the DD model.  It is also worth noting that in the
absence of a driving field, the dynamics again reduces to a DD process, but
with a length-dependent hopping rate that is proportional to the inverse
domain length \cite{cor,majhuse}.

\begin{figure}
\narrowtext
\epsfxsize=6cm\epsfysize=3.75cm\hskip 1.5cm
\epsfbox{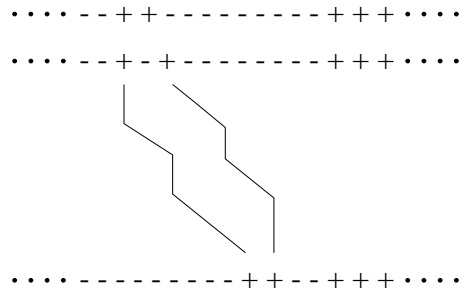}
\vskip 0.15in
\caption{Time evolution of an up dimer in a sea of down spins.  When the dimer
  dissociates, the isolated spins independently hop to the right.  Shown is
  the space-time trajectory of the dimer for the case where the dimer
  recombines and becomes stationary again before the next domain wall is
  reached.
\label{fig2}}
\end{figure}

A crucial feature of the mapping between the spin and the domain dynamics
which is not apparent from the above description is the evolution of dimers
(see Fig.~2).  Consider an up dimer.  If the rightmost spin of the dimer
detaches, the dimer is converted into two isolated up spins in a sea of down
spins.  According to the spin dynamics, each isolated spin independently and
freely hops to the right.  Consequently their separation undergoes a simple
random walk.  The motion of spins anti-parallel to the field can be
neglected, since this motion is inhibited by a factor $e^{-2E/T}$.  The
hopping of this pair of separated spins terminates in one of two ways: (i)
The rightmost up spin reaches the next domain boundary and subsequently the
other up spin hits this same boundary.  This corresponds to the coalescence
of the two adjacent down domains and is part of the DD picture.  (ii) The
dimer recombines {\em before} the next domain boundary is reached
(Fig.~\ref{fig2}).

Dimer recombination is the crucial new feature which was not included in the
DD process.  This recombination plays an essential asymptotic role because
the average dimer jump distance $\ell$ is much smaller than the average
domain length $L$ in the long-time limit.  Consequently, recombination of a
dimer is much more probable than domain coalescence.  To verify this
assertion let us estimate $\ell$.  The dimer recombines if the separation
between the two spins shrinks to zero before the rightmost spin reaches the
next domain boundary.  This is a classic first-passage process, and the
probability that this separation first equals zero at $2l$ steps is given by
\cite{feller}
\begin{equation}
\label{Pl}
{\cal P}(l)={1\over{2^{2l-1}}}{(2l-2)!\over{(l-1)!\,l!}}
\end{equation} 
For large $l$, this expression simplifies to ${\cal P}(l)\sim \pi^{-1/2}
l^{-3/2}$.  The average dimer jump distance $\ell$ may now be estimated as
$\ell=\sum_{l\leq L} l\,{\cal P}(l)\sim \sqrt{L}$.  Thus asymptotically
$\ell\ll L$.

We now use this picture to estimate the overall time dependence of
small-length domains.  The crucial feature is that a domain can disappear
only if a dimer first dissociates and then does not recombine before its
constituent spins reach the next domain boundary.  From the analogy with the
first-passage process, the probability $R(t)$ that the dissociated dimer does
not recombine before the next domain boundary is given by
\begin{equation}
\label{trap}
R\sim \sum_{l=L}^{\infty} {\cal P}(l)\sim \sum_{l=L}^\infty l^{-3/2} \sim L^{-1/2}.
\end{equation}
Since the disappearance of a dimer leads to domain coalescence, the total
number of domains $N(t)$ obeys the rate equation
\begin{equation}
\label{Ndot}
\frac{dN}{dt}= -RC_2 \sim -{C_2\over \sqrt{L}}.
\end{equation}

On the other hand, the dynamics of large domains should still be governed by
the gain and loss of single spins at the boundary, as outlined in
Fig.~\ref{fig1}.  Since these gain and loss processes occur at the same rate,
the domain length undergoes an isotropic random walk, so that $L(t)$ should
grow as $t^{1/2}$.  Correspondingly, the number of domains $N(t)$ decays as
$t^{-1/2}$, the inverse of the average domain length.  Substituting these two
expectations into Eq.~(\ref{Ndot}), we immediately obtain
\begin{equation}
\label{C2}
C_2(t)\sim t^{-5/4}.
\end{equation}
This slower decay of dimers is one of the primary features of the DC process.
It should be compared with the prediction $C_2(t)\sim t^{-3/2}$ which arises
from the DD process.  This latter time dependence would be obtained from
Eq.~(\ref{Ndot}) if the rate of dimer disappearance were unity, rather than
proportional to $L^{-1/2}$.

This $-5/4$ exponent value has fundamental implications for the density of
domains of length $k$, $C_k(t)$.  Let us suppose that this density obeys the
conventional scaling hypothesis
\begin{equation}
\label{scaling}
C_k(t)\sim {1\over L^2}F\left({k\over L}\right),
\end{equation}
where the prefactor $L^{-2}$ follows from the length normalization condition
$\sum k\,C_k(t)=1$.  If $C_2(t)\sim t^{-5/4}$, then either $F(z)\sim
\sqrt{z}$ as $z\to 0$, or $C_k(t)$ does not obey scaling for small $k$.  We
shall present evidence from both simulations and an analytical approach that
strongly favors the latter alternative.

\section{Simulation results}

In our simulations, we first initialize an array of alternating up and down
domains of random lengths.  For minority phase density $\mu$, we choose the
average length of minority domains to be $L=10$, and $L/\mu$ for the majority
phase.  The time evolution involves the following steps: (i) Pick a domain at
random.  (ii) Move an up (down) domain of length $>2$ to the right (left).
(iii) If the domain is a dimer, choose its jump distance $l$ from the
probability distribution ${\cal P}(l)$ given by Eq.~(\ref{Pl}).  If $l$
exceeds the length of the neighboring domain, remove the dimer and merge the
surrounding domains.  (iv) Update the time by $1/({\rm number\ of\ 
  domains})$.  Simulations were performed on a chain of $4\times 10^6$
domains for times up to $5\times 10^5$, and averaged over $16$ samples.  This
is of the same order of data as the simulations of Cornell and
Bray\cite{cobra}.

\subsection{Average time-dependent properties}

\begin{figure}
\narrowtext
\epsfxsize=7.5cm\epsfysize=7.5cm
\hskip 0.0in\epsfbox{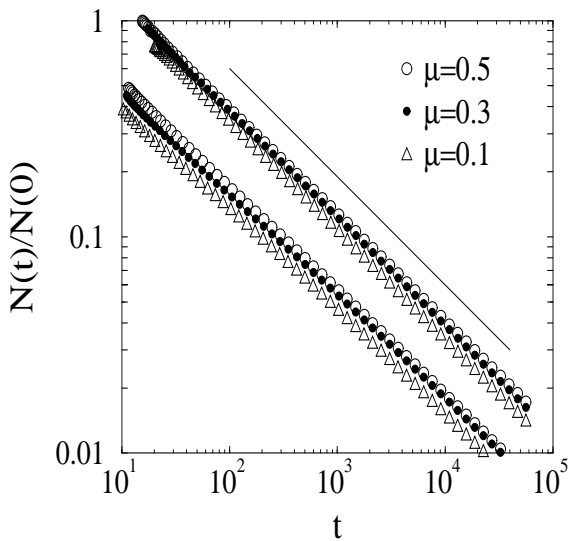}
\vskip 0.15in
\caption{Time dependence of the domain density for various minority spin fractions 
  $\mu$.  Upper set of points -- DC data, lower set -- DD data.  The DD data
  are divided by $2$ to separate the two sets.  As a guide to the eye, the
  solid line has a slope $-1/2$.
\label{nvst}}
\end{figure}

\begin{figure}
\narrowtext
\epsfxsize=7.5cm\epsfysize=7.5cm
\hskip 0.0in\epsfbox{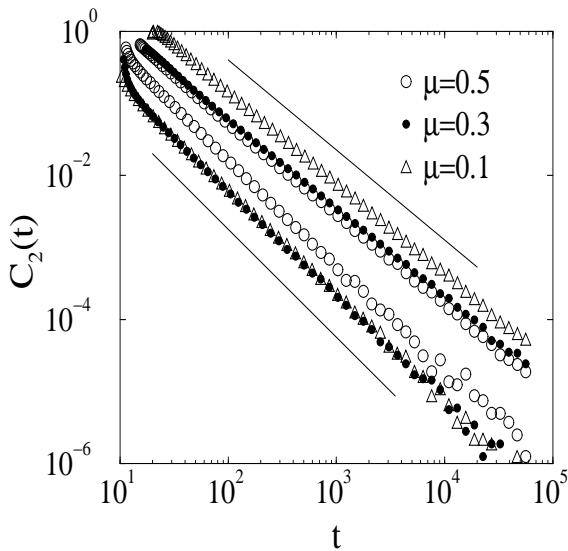}
\vskip 0.15in
\caption{The density of dimers as a function of time.  Upper points -- DC,
  lower points -- DD.  As a guide to the eye, the solid lines have slopes
  $-5/4$ (upper) and $-3/2$ (lower).
\label{ctwovst}}
\end{figure}

In Fig.~\ref{nvst}, we plot $N(t)$ for various minority fractions $\mu$.
Also shown are the corresponding results for the DD process.  The linearity
of the data suggests power-law behavior, and visually the asymptotic slopes
are very close to the expected value of $-1/2$.  The essential difference
between DC and DD models is manifested by the behavior of the dimer density.
Fig.~\ref{ctwovst} shows that the dimer density indeed decays more slowly
than in the DD model.

\begin{figure}
\narrowtext
\epsfxsize=7.5cm\epsfysize=7.5cm
\hskip 0.0in\epsfbox{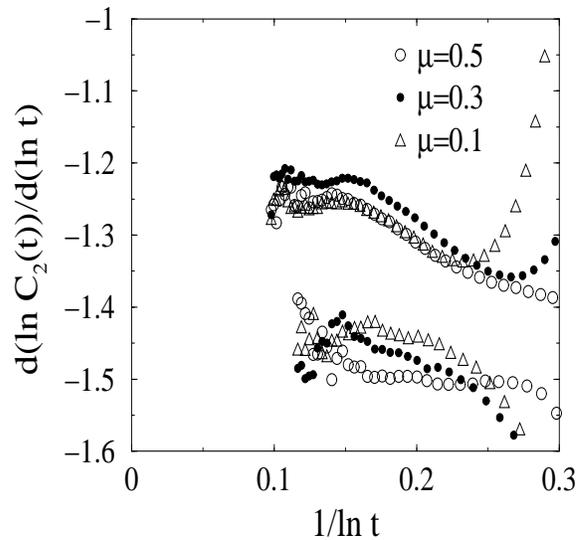}
\vskip 0.15in
\caption{Local exponents for the time dependence of the number of dimers.  
  Upper points -- DC model, lower points -- DD model.  
\label{locexp1}}
\end{figure}

To highlight this difference, we plot the corresponding local exponents in
Fig.~\ref{locexp1}.  We define the local exponent at time $t$ as the best-fit
straight line to 10 successive data points (equally spaced on a logarithmic
scale) up to time $t$ in the double logarithmic plot of $C_2(t)$ versus $t$.
This definition significantly smooths statistical fluctuations while still
revealing systematic trends in the data.  As shown in Fig.~\ref{locexp1}, the
local exponents of the DD and DC models are clearly different.  For the DD
model, these exponents are close to the expected value $-3/2$ and also
exhibit weak systematic time dependence.  Thus, the natural conclusion is
that the dimer density (as well as the density of domains of any fixed
length) asymptotically decays as $t^{-3/2}$.

\begin{figure}
\narrowtext
\epsfxsize=7.5cm\epsfysize=7.5cm
\hskip 0.0in\epsfbox{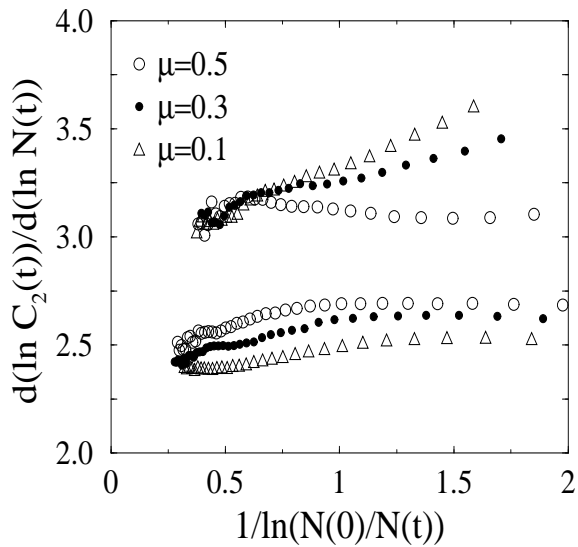}
\vskip 0.15in
\caption{Local exponents for the dependence of the number
  of dimers on the total number of domains.  Lower points -- DC model, upper
  points -- DD model.
\label{cvsn}}
\end{figure}

For the DC model, the situation is more subtle.  The local exponents
initially are increasing with time, but this time dependence slows when the
effective exponent value is close to the anticipated value $-5/4$.  However,
the systematic ambiguities in the data make an extrapolation for the
asymptotic value of the exponent uncertain.  This uncertainty and the
relatively small difference in the dimer exponent for the two models led us
to consider $C_2(N)$ rather than $C_2(t)$.  Indeed, since both $N(t)$ and
$C_2(t)$ should be influenced by the same pre-asymptotic corrections, such
corrections might cancel when $C_2$ is expressed as a function of $N$.  From
a scaling perspective, it is also natural to express dependences in terms of
intrinsic variables, rather than in terms of the extrinsic time variable.
Thus in Fig.~\ref{cvsn}, we plot the local exponents of $C_2(N)$ versus $N$
and the results are now relatively straightforward to interpret.  Since the
variation in the local exponent for the DC model is small over the entire
range of $N$, the result $C_2(N)\sim N^{5/2}$ is strongly suggested.
Similarly, the DD data suggests that $C_2(N)\sim N^3$, as is anticipated from
$N(t)\sim t^{-1/2}$ and $C_2(t)\sim t^{-3/2}$.  Coupled with the basic result
$N(t)\sim t^{-1/2}$ which holds for both models, we conclude, now with
considerable confidence, that $C_2(t)\sim t^{-5/4}$ for the DC model.

\subsection{Domain length distribution}

\begin{figure}
\narrowtext
\epsfxsize=7.5cm\epsfysize=7.5cm
\hskip 0.0in\epsfbox{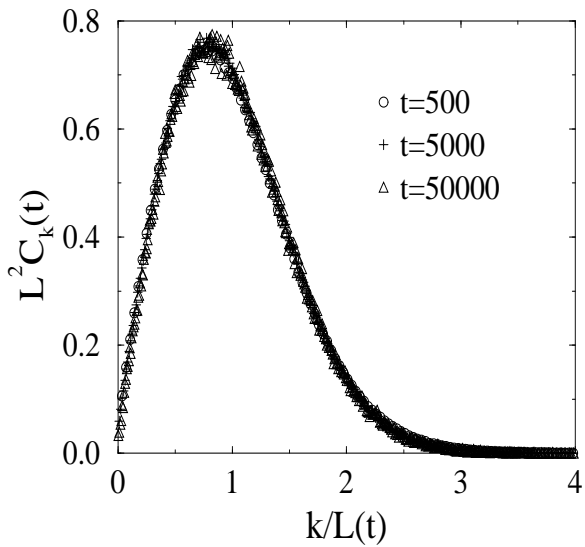}
\vskip 0.15in
\caption{Scaling plot for the domain-length distribution for 
  equal fractions of up and down spins.  Notice that this distribution does
  not reach zero at $k/L=0$ (see Fig.~\ref{summ3}).
\label{summ2}}
\end{figure}

The behavior of the domain length distribution is especially interesting, as
scaling breaks down in the small-length limit.  This ultimately arises from
the multi-step dissociation and recombination processes that govern the
disappearance of dimers.  We first test the conventional scaling hypothesis
for the domain length distribution, namely, $C_k(t) \sim L^{-2} F(k/L)$, by
plotting $L^2C_k(t)$ versus $k/L$ in Fig.~\ref{summ2}.

\begin{figure}
\narrowtext
\epsfxsize=7.5cm\epsfysize=7.5cm
\hskip 0.0in\epsfbox{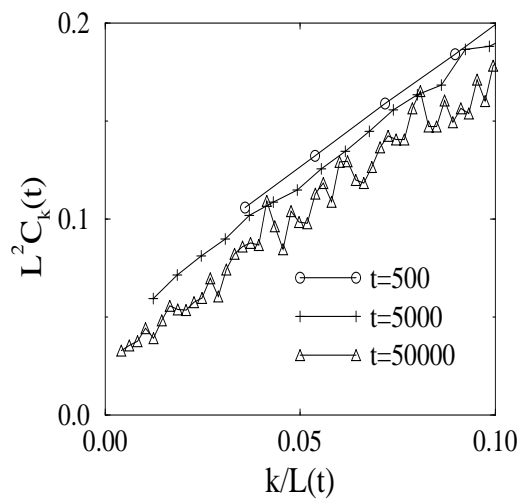}
\vskip 0.15in
\caption{Small-length tail of the domain length distribution 
  for equal fractions of up and down spins, showing the systematic departure
  from scaling and an intercept at $k=0$ which is systematically decreasing
  with time.
\label{summ3}}
\end{figure}

At the scale shown, this distribution appears to exhibit data collapse.  In
fact, at the resolution of this figure, the domain length distributions for
the DD and DC processes are virtually indistinguishable.  However, for the
scaling form to be compatible with $C_2(t)\sim t^{-5/4}$, the scaling
function must vary as $F(z)\sim z^{1/2}$ as $z\to 0$, while the length
distribution appears to be linear in $k$ in the small-length limit.  This
linearity implies that the distribution cannot obey single-parameter scaling
for all lengths.  In fact, a closer examination of the small-$k$ tail
(Fig.~\ref{summ3}) reveals that there is that there is a small but systematic
deviation from data collapse.  The breakdown of scaling is manifested by the
length distribution having a non-zero intercept with the $k=0$ axis, whose
value is systematically decreasing with time.

\section{Domain length distribution in the limit $\mu\rightarrow 0$}

To better understand the nature of the domain length distribution, we focus
on the limit where the fraction of minority spins $\mu$ is vanishingly small.
This leads to considerable simplification in the domain dynamics.  Generally,
the length of a domain can change by $\pm 1$ due to diffusion of neighboring
domains, or the length can change arbitrarily by domain coalescence.  In the
limit $\mu\to 0$, the diffusive ``shrinkage'' governs the disappearance of
minority domains, while coalescence governs the disappearance majority
domains.  To verify this, let us estimate the characteristic times for the
disappearance of the majority and minority domains by shrinkage.  Let $L_-$
($L_+$) denote the average length of minority (majority) domains.  A majority
domain can shrink to zero in a time $t_+$ of order $L_+^2$, while the
shrinking and disappearance of a minority domain requires a time $t_-\sim
L_-^2$.  Thus
\begin{equation}
\label{tpm}
{t_-\over t_+}\sim \left({L_-\over L_+}\right)^2=\mu^2.
\end{equation}
Therefore in the minority limit, shrinkage of majority domains, or
equivalently, coalescence of minority domains is negligible.  Consequently
the minority domains are effectively non-interacting and they evolve only by
the addition or loss of single spins as a result of the hopping of majority
domains.  Therefore the density of minority domains $C_k(t)$ obeys the
discrete diffusion equation
\begin{equation}
{dC_k\over dt}=C_{k+1}-2C_k+C_{k-1}, \quad k>2.
\label{ck}
\end{equation}

The density of dimers $(k=2)$ obeys a separate, but similar equation.  For
dimers, there is no gain term due to processes which involve monomers, and
the loss of dimers due to their dissociation into two monomers and ultimate
domain coalescence occurs at a rate $R\sim L^{-1/2}$ (see Sec.\ II), since a
dissociated dimer may recombine before the coalescence occurs.  Therefore the
master equation for $C_2(t)$ is
\begin{equation}
{dC_2\over dt}=C_3-C_2-{C_2\over \sqrt{L}}. 
\label{c2}
\end{equation}

In the continuum limit, Eq.~(\ref{ck}) is equivalent to 
\begin{equation}
{\partial C_k(t)\over \partial t}={\partial^2 C_k(t)\over\partial k^2},
\label{diff}
\end{equation}
while Eq.~(\ref{c2}) provides the boundary condition.  In this equation, the
left-hand side scales as $t^{-1}C_2$, while the last term on right-hand side
scales as $L^{-1/2}C_2\sim t^{-1/4}C_2$.  Thus in the long-time limit the
left-hand side is negligible and Eq.~(\ref{c2}) becomes
\begin{equation}
\left[{\partial C_k(t)\over \partial k}-{C_k(t)\over \sqrt{L}}\right]
_{k\rightarrow 0}=0.
\label{bound}
\end{equation}
This radiation boundary condition\cite{weiss} expresses the fact that a dimer
does not necessarily disappear when it dissociates, but it may be
reconstituted and then grow into a finite-size domain.

By dimensional analysis, Eq.~(\ref{diff}) implies the existence of the usual
diffusive length scale $L=\sqrt{t}$.  By similar reasoning, Eq.~(\ref{bound})
suggests the existence of an additional length scale $\ell=\sqrt{L}$.  The
competition between these two scales determines the asymptotic behavior.  We
therefore separately consider the ``inner'' region of small domains $k\ll L$
and the ``outer'' region of large domains $k\gg \ell$, and then match these
limiting solutions in the overlap region $\ell\ll k\ll L$ \cite{nayfeh}.

In the inner region $k\ll L$, the diffusion equation simplifies to
${\partial^2 C\over \partial k^2}=0$, whose solution is $C_k(t)=A(t)+B(t)k$.
Employing the boundary condition Eq.~(\ref{bound}) we obtain
\begin{equation}
C_k(t)^{\rm inner}=A(t)\left(1+{k\over\sqrt{L}}\right).
\label{inner}
\end{equation}
In the outer region $k\gg \ell$, the system is governed by the original
diffusion equation (\ref{diff}), while Eq.~(\ref{bound}) reduces to the
absorbing boundary condition.  The solution in this region thus becomes
\begin{equation}
C_k(t)^{\rm outer}={k\over t^{3/2}}~\exp\left(-{k^2\over 4t}\right).
\label{outer}
\end{equation}
The inner and outer solutions should match in the overlapping region $\ell\ll
k\ll L$.  This determines the amplitude $A(t)$ in Eq.~(\ref{inner}) to be
proportional to $t^{-5/4}$.   The inner solution now becomes
\begin{equation}
C_k(t)^{\rm inner}={\gamma\over t^{5/4}}+{k\over t^{3/2}}
\label{in}
\end{equation}
with $\gamma$ a constant.  

\begin{figure}
\narrowtext
\epsfxsize=7.5cm\epsfysize=7.5cm
\hskip 0.0in\epsfbox{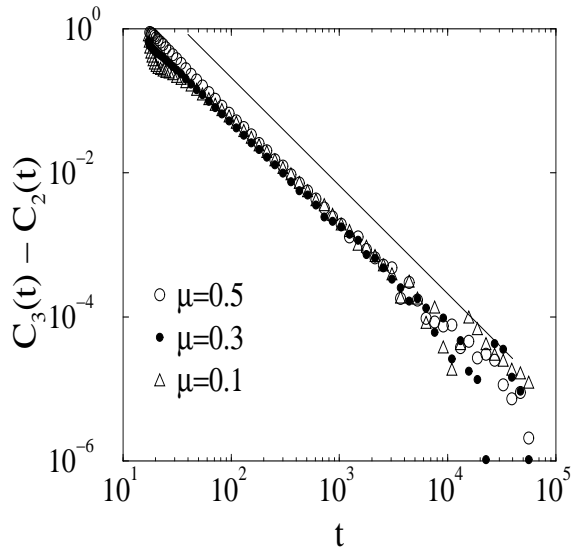}
\vskip 0.15in
\caption{Plot of $C_3(t)-C_2(t)$ versus time.  As a guide to the eye, a straight 
  line of slope $-3/2$ is shown.
\label{correction}}
\end{figure}

These two limiting forms for $C_k(t)$ match smoothly in the overlap region
$\ell\ll k\ll L$.  This further suggests that the domain length distribution
for the entire length range can be accounted for by the composite form
\begin{equation}
\label{comp}
C_k(t)=\left({\gamma\over t^{5/4}}+{k\over t^{3/2}}\right)\,
\exp\left(-{k^2\over 4t}\right).
\end{equation}
To determine the validity of this hypotheses, we test for the existence of
the $k/t^{3/2}$ correction term, since the leading $t^{-5/4}$ time dependence
has already been established.  For this purpose, consider $C_3(t)-C_2(t)$
versus $t$.  This difference eliminates the leading $t^{-5/4}$ behavior and
thus isolates the $k/t^{3/2}$ correction term (Fig.~\ref{correction}).  As
seen in the figure, the data for $C_3-C_2$ is consistent with a $t^{-3/2}$
time dependence.  This test also supports the correctness of the composite
form of Eq.~(\ref{comp}) for the domain length distribution.  Finally, our
numerical data suggests that this same dependence holds for all values of
$\mu$.

\section{Summary and Discussion}
 
We investigated the low-temperature coarsening of an Ising chain subject to
spin-exchange dynamics and a weak driving force.  The spin dynamics was
reduced to a dimer-controlled domain diffusion process.  From this picture,
we established the existence of {\em two} growing characteristic length
scales; one is the fundamental diffusion length $t^{1/2}$, which provides the
average domain size, and the other, $\ell\sim t^{1/4}$, is the average dimer
hopping distance.  The competition between these two scales leads to a
breakdown of conventional scaling in the small-length tail of the domain
distribution.  As a consequence, the density of fixed-length domains decays
as $t^{-5/4}$ as $t\to\infty$.  A key step to verify this latter result was
to study the dependence of $C_k$ on $N$, rather than the dependence on $t$.

For the one-dimensional system, several basic unresolved issues remain.  Thus
far, an analytical solution for the length distribution of minority domains
only has been obtained in the extreme minority limit.  It would be worthwhile
to study analytically the case of arbitrary minority fraction.  One approach
is to treat domains as statistically independent.  Such an approximation is
exact in the extreme minority limit and also works well for the coarsening of
the undriven Ising chain for both spin-flip and spin-exchange
dynamics\cite{ab,dz,ep}.  Under the assumption of statistical independence,
it is possible to solve the rate equations for $C_k(t)$.  This solution
reproduces the correct dynamical exponent, as well as the linear small-length
tail for the domain length distribution\cite{noiseless}.  This approach
further predicts an exponential decay for the the large-length limit of the
domain length distribution, for any non-zero fraction of minority spins.
However, our numerical simulations at zero magnetization suggest that this
large-length tail has the leading behavior $\exp(-(k/L)^\nu)$, with $\nu$
greater than 1 and less than 2.  This puzzling feature, neither an
exponential nor a Gaussian decay for the large-length tail of the
distribution, deserves more careful attention.

\medskip
We gratefully acknowledge partial support from NSF grant DMR9632059 and ARO
grant DAAH04-96-1-0114.

\end{multicols}
\end{document}